\documentclass[aoas,nameyear,dvips]{arximspdf}

\doi{10.1214/10-AOAS386}
\volume{4}
\issue{4}
\pubyear{2010}
\firstpage{1657}
\lastpage{1659}

\begin{document}
\begin{frontmatter}

\title{Remembering Leo}
\runtitle{Remembering Leo}

\begin{aug}
\author{\fnms{Bin} \snm{Yu}\corref{}\ead[label=e1]{binyu@stat.berkeley.edu}}
\runauthor{B. Yu}
\affiliation{University of California, Berkeley}
\address{Departments of Statistics and EECS\\
 University of California, Berkeley\\
 Berkeley, California 94720\\
USA\\
\printead{e1}} 
\end{aug}

\received{\smonth{7} \syear{2010}}



\end{frontmatter}

I do not remember when was the first time that I met Leo, but
I have a clear memory of going to Leo's office on the 4th floor
of Evans Hall to talk to him in my second year
in Berkeley's Ph.D. program in 1986. The details of the
conversation are not retained but a visual image of his clean
and orderly office remains, in a stark contrast to a high entropy
state of the same office now being used by myself.

In the 1980s, the Berkeley Statistics Department's Statistical Computing Facility (SCF) managed
mainframe computers named after musicians such as Bizet and Bach.
Later I learned that Leo came back
to Berkeley in 1980 as a professor from industrial consulting
and soon established and directed SCF---a far-sighted move by any measure,
and one that benefited the department for decades to come.
Before his retirement in January 1993,
Leo had not only been an energetic and effective director of SCF
but also an energetic and effective teacher in his multivariate
statistics graduate class, in addition to much original research.
After his retirement,
he remained a strong leader of statistical computing in the department
by advising students and helping start the statistical machine
learning group with many joint appointments with Computer Science and
Electrical Engineering.
Moreover, as he told me himself, this period
after retirement was the best time for his research---he was producing seminal papers one after the other
on statistical machine learning, for example, on analyzing
boosting and developing now-popular algorithms such as bagging and random forests.

After I returned to Berkeley as a faculty member in 1993, I would frequently
encounter Leo on the 4th floor of Evans Hall since my first office was also
on the same floor. We would have hallway conversations and he would
tell me what was on his mind at that time.
During one of those conversations, Leo asked me
to join a journal club on neural networks, one of the
topics that attracted Leo's attention at that time and the beginning
of his work on understanding machine learning methods and
developing many of his own. I believe Jerry Feldman from CS and ICSI
was jointly organizing the journal club. The time must have been
around 1995 or 1996 because during that time Leo ``dragged'' me
to Denver for the NIPS in 1996 for the first time and that was the only time
that I have attended NIPS so far (although I have been co-authoring NIPS
papers lately). Those\vadjust{\goodbreak} interactions with Leo in the late 1990s planted the seeds for my
interest and research in statistical machine learning.

In the neural network journal club, I was introduced
for the first time to the early stopping idea of regularization which was
also used in boosting later and this idea has fascinated me for the past
many years.
I heard of AdaBoost for the first time also from Leo in one
of those hallway conversations. Leo was convinced of the empirical
success of AdaBoost, but was very puzzled by its mysterious
``overfitting resistance'' property. Leo's interest in understanding why
AdaBoost works might have led him to formulate or rederive AdaBoost
from an optimization point of view: AdaBoost could be viewed
as gradient descent on an exponential loss function of the margin [\citet{Breiman1999}]. Building on this work of Leo
and the paper by Friedman, Hastie and Tibshirani (\citeyear{FHT2000}) on boosting, Buhlmann and I used L2Boosting
to explain to a certain extent the mysterious ``overfitting resistance'' [\citet{BuhlmannYu2003}].

Leo was honored in 2002 with the prestigious Wald Lecturer by IMS and
I was asked to chair his first Wald Lecture given at JSM.
It was a great opportunity for me to discover and present Leo's other dimensions
rather than academic
in my short introduction to his lecture. Several slides were used
and they included the cover of the SIAM reprint of Leo's celebrated
book {\it Probability}, Leo's sculptures, his house that
he had helped design, and maybe even a photo related to his days working
on school boards at LA.

I was fortunate to see Leo in the spring of 2005, when he was already very weak and resting at home---he seemed very pleased to see me.
The news of his passing came to me
when I was teaching a short summer course at Peking University on information theory and statistics, a topic to which Leo made significant contributions
earlier in his career, including the Shannon--McMillan--Breiman theorem
in information theory. This famous theorem
shows the convergence of the average negative log probability of an $n$-tuple to the entropy of the process
for discrete-time finite-valued stationary ergodic sources.
I stopped my lecture and told the students about Leo.

Leo was and still is one of my heros.  His paper on predictive
statistics [\citet{Breiman2001}] is a must-read in my graduate applied statistics class
at Berkeley. And I am
inspired every time when I think of Leo's creativity, originality, and humanity.

\printaddresses

\end{document}